\newif\ifmulticol	\multicoltrue
\newif\ifshowgit	\showgittrue		
\newif\ifgitlocal	\gitlocaltrue		
\newif\ifbiblatex	\biblatexfalse		
\newif\ifbibnum		\bibnumtrue 		
\newif\ifbibsort	\bibsortfalse		
\newif\iflineno		\linenofalse
\newif\iftoc		\tocfalse

\newif\iflucida		\lucidafalse
\newif\ifcm			\cmfalse
\newif\iflibertine	\libertinefalse		
\newif\ifcharter	\chartertrue

\newcommand*{\secnumdepth}{0}			
\newcommand*{\tocdepth}{2}				
\newcommand*{\reftitle}{References}		
\newcommand*{\mytitleskip}{-0.2in}		

\newcommand*{\mydocfontsize}{\ifcharter11pt\else\iflibertine11pt\else10pt\fi\fi}
\newcommand*{\setcol}{\ifmulticol twocolumn\else onecolumn\fi}

\documentclass[\mydocfontsize,\setcol]{article}

\newcommand*{\pdftitle}{Disease from opposing forces in regulatory control}

\input oppose.sty

\usepackage{bm}

\DeclarePairedDelimiter\abs{\lvert}{\rvert}
\DeclarePairedDelimiter\norm{\lVert}{\rVert}
\DeclarePairedDelimiter\angb{\langle}{\rangle}
\DeclarePairedDelimiter\lrb{\lbrack}{\rbrack}
\DeclarePairedDelimiter\lr{\lparen}{\rparen}
\DeclarePairedDelimiter\lrbr{\lbrace}{\rbrace}

\makeatletter
\let\oldabs\abs \def\abs{\@ifstar{\oldabs}{\oldabs*}}
\let\oldnorm\norm \def\norm{\@ifstar{\oldnorm}{\oldnorm*}}
\let\oldangb\angb \def\angb{\@ifstar{\oldangb}{\oldangb*}}
\let\oldlrb\lrb \def\lrb{\@ifstar{\oldlrb}{\oldlrb*}}
\let\oldlr\lr \def\lr{\@ifstar{\oldlr}{\oldlr*}}
\let\oldlrbr\lrbr \def\lrbr{\@ifstar{\oldlrbr}{\oldlrbr*}}
\makeatother

\newcommand*{\Figure}[1]{Figure~\ref{fig:#1}}
\newcommand*{\Fig}[1]{Fig.~\ref{fig:#1}}

\newcommand*{\boldrule}{\hrule height 1.2pt}
\newcommand*{\noterule}{\medskip\boldrule\medskip}	

\newcommand*{\reviset}[1]{\textcolor{black}{#1}}

\newcommand*{\mytitle}{\pdftitle}

\newcommand*{\myauthor}{Steven A.\ Frank}
\newcommand*{\myaddress}{Department of Ecology and Evolutionary Biology, University of California, Irvine, CA 92697--2525, USA}
\newcommand*{\mycontact}{\href{https://stevefrank.org}{https://stevefrank.org}}

\newcommand*{\mykeywords}{Genomic imprinting, IGF2, sexual antagonism, genomic conflict, immune system disorder, dopamine, psychiatric disorder\\ \\ \textbf{Page heading:} Disease from opposing forces}

\setabstract{-0.07}{\iftoc-2.0\else1.22\fi}{%
Danger requires a strong rapid response. Speedy triggers are prone to false signals. False alarms can be costly, requiring strong negative regulators to oppose the initial triggers. Strongly opposed forces can easily be perturbed, leading to imbalance and disease. For example, immunity and fear response balance strong rapid triggers against widespread slow negative regulators. Diseases of immunity and behavior arise from imbalance. A different opposition of forces occurs in mammalian growth, which balances strong paternally expressed accelerators against maternally expressed suppressors. Diseases of overgrowth or undergrowth arise from imbalance. Other examples of opposing forces and disease include control of dopamine expression and male versus female favored traits.
}

\begin{document}

\mymaketitle

\iftoc\mytoc{-24pt}{\newpage}\fi

\section{Introduction}

Strongly opposing forces on traits may lead to imbalance and disease \autocite{moore91genomic}. For example, paternally expressed genes such as \textit{IGF2} accelerate mammalian growth. Maternally expressed genes such as \textit{H19} slow growth. The opposition of strong growth promoters and suppressors creates a precarious balance. Disruption of growth promoters causes the suppressors to dominate, leading to deleterious undergrowth. Perturbation of suppressors leads to deleterious overgrowth \autocite{fowden11imprinted}.

To develop the general argument, I start with this example of mammalian growth. The opposing forces have been identified. Disruption of those forces leads to known examples of disease. Using that growth example as a model, I then summarize two previously discussed candidates.

First, X chromosomes and autosomes conflict over traits that have different favored expression levels in males and females. Some parts of the genome tend to push expression toward the male-favored traits. Other parts of the genome tend to push expression toward the female-favored traits. Perturbation may lead to imbalance and disease along a male-female trait axis \autocite{frank11pathology}.

Second, immune expression is often triggered very rapidly to control infection. The need for a speedy response may lead to mistakes caused by false triggers of immunity. To suppress false immune triggering, strong negative regulators of immunity oppose the initial triggers. Perturbation of these opposing forces on immune expression may lead to misregulation and disease \autocite{frank19evolution}.

Finally, I add a new candidate. A speedy response to potential danger often associates with fear. The rapid fear response may be prone to false signals. Powerful negative regulators oppose false triggers to maintain balance. Perturbation of those opposing forces may lead to misexpression and various psychiatric disorders.

\begin{figure*}[t]
\centering
\includegraphics[width=250pt]{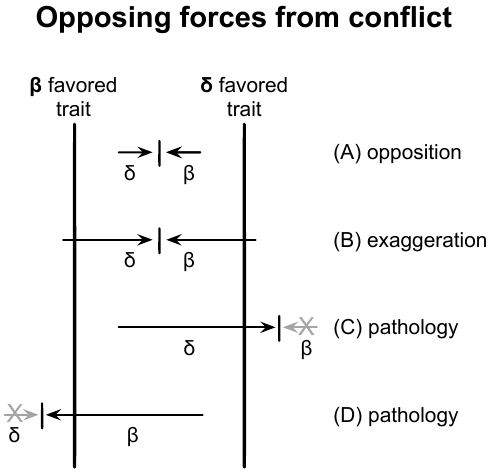}
\caption{\reviset{One way in which opposing forces from conflict may lead to disease. Two conflicting parties, $\beta$ and $\delta$, favor different trait values. (A) The conflicting parties develop opposing forces acting on the trait. (B) Over time, the conflict favors stronger opposing forces, leading to exaggeration. (C) With strongly opposing forces, any perturbation in the delicate balance may lead to dominance by one party and overexpression of the trait past its favored value, causing disease. In this case, the ``X'' represents a knockout of one side in the conflict or a mismatch in the mediation of the conflict that leads to dominance by one party and imbalance in trait expression. (D) Similarly when the alternative party dominates. The example in the text of genetic conflict over mammalian growth provides one well-established case. The example of male-female conflict described in the text may be another case. Other cases arising from genomic conflicts have been described \autocite{frank89the-evolutionary,frank91divergence,hurst91causes,johnson10hybrid}. This figure evokes only a very rough intuitive sense of the idea and is not meant to be interpreted precisely.}}
\label{fig:conflict}
\end{figure*}

\begin{figure*}[t]
\centering
\includegraphics[width=250pt]{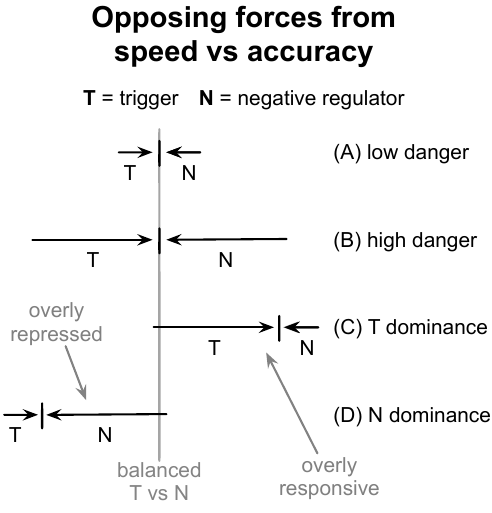}
\caption{\reviset{One way in which the speed versus accuracy tradeoff may lead to opposing forces and disease. A fast trigger (T) responds a potential signal of danger. Negative regulators (N) oppose the trigger to correct false signals. (A) Low danger frequency favors relatively slow and weak triggers and negative regulators. (B) High danger frequency favors fast and strong triggers with a higher rate of false alarms, which leads to faster and stronger negative regulators. (C) Individuals that have strong triggers and weak negative regulators tend to be overly responsive, potentially causing pathological response. (D) Individual that have weak triggers and strong negative regulators tend to be overly repressed, potentially causing pathological response. Mismatch between triggers and negative regulators may arise by mutation, regulatory perturbation, or mating between parents with genes that are tuned to different levels of environmental danger. This figure evokes only a very rough intuitive sense of the idea and is not meant to be interpreted precisely.}}
\label{fig:speed}
\end{figure*}

\reviset{In the mammalian growth and X-autosome antagonism examples, the opposing forces arise from conflict (\Fig{conflict}). In the immune and danger responses, the opposing forces arise from the tradeoff between the speed and accuracy of the response (\Fig{speed}).}

\reviset{Others have noted how misregulation and disease may arise in speed versus accuracy tradeoffs. For example, Nesse's smoke detector principle describes how the need for a speedy response leads to a significant frequency of false signals \autocite{nesse01the-smoke,nesse19the-smoke}. However, that signal-detection theory does not focus on the natural tendency for regulatory mismatch between strongly opposing forces and the consequences for an elevated rate of disease. The value here arises from the broad emphasis on that regulatory mismatch of opposing forces as a general principle in the study of disease.}

\begin{figure*}[t]
\centering
\includegraphics[width=0.55\hsize]{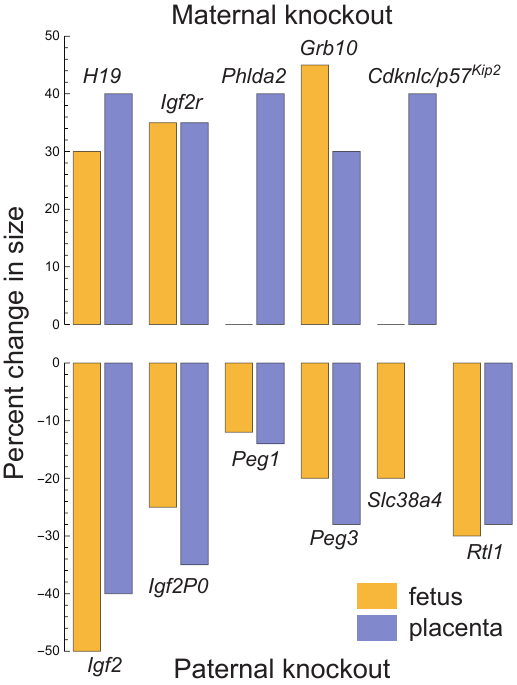}
\caption{Knockouts of paternally expressed growth promoters cause severe undergrowth in mice. Knockouts of maternally expressed growth suppressors cause large overgrowth. The bottom shows the percent decrease in size of the fetus and placenta for each single independent knockout of six different paternally expressed growth promoter genes. Similarly, the top shows the size increase for single independent knockouts of five different maternally expressed growth suppressor genes. The placental measure for \textit{Slc38a4} is missing. Data from Table 1 of Fowden et al. \autocite{fowden11imprinted}.}
\label{fig:imprint}
\end{figure*}

\section{Mammalian growth}

In mammals, a diploid genetic locus derives from one paternally inherited allele and one maternally inherited allele. Many growth regulating loci express one parent's allele and silence the other. Typically, growth promoting genes express the paternal allele. Growth suppressing genes express the maternal allele. Growth ultimately depends on these opposing forces \autocite{fowden11imprinted}.

\Figure{imprint} shows that the growth promoters and suppressors strongly oppose each other. Knockout of any single paternally expressed promoter greatly reduces the size of the fetus and placenta. In most cases, knockout of any maternally expressed suppressor greatly increases size. Growth appears to be a precarious balance between strongly opposed forces.

Moore and Haig suggested that the opposed forces may arise from the conflicting interests of fathers and mothers on offspring growth rate \autocite{moore91genomic}. That conflict creates an evolutionary tug-of-war that leads to the strongly opposing forces on growth. Such precarious balance is easily perturbed, leading to disease.

For example, higher expression than normal of paternally expressed \textit{IGF2} or lower expression than normal of maternally expressed \textit{H19} or \textit{CDKN1C} leads to a broad spectrum of overly rapid growth pathologies known as Beckwith–Weidemann syndrome. By contrast, Silver-Russell syndrome is an undergrowth pathology that sometimes associates with opposite changes in growth suppressors or promoters. 

Moore and Haig's idea that strongly opposing forces predispose to disease leads to the following examples.

\section{Male-female trait conflict}

Males and females often have different favored values for traits \autocite{rice01intersexual}. Selection in males pushes the trait in one direction, and selection in females pushes the trait in the other direction. An autosomal genetic locus is symmetric with respect to male and female interests. Thus, the opposing male and female selective pressures tend to balance, leading to an intermediate phenotype.

The situation differs on the X chromosome of species with XX females and XY or XO males \autocite{rice84sex-chromosomes}. Depending on particular genetic assumptions about dominance and dosage effects, X-linked loci may tend to push more strongly toward the female or male optimum than do autosomal loci \autocite{frank20sexual}. Because conflict occurs between different chromosomes, the opposing forces tend to increase. Any additional push by the autosomes toward an intermediate phenotype may be opposed by an additional X-linked push toward the favored sex. 

This fundamental conflict likely leads to strongly opposed regulatory forces \autocite{frank11pathology}. As those strongly opposed forces arise, susceptibility to perturbation and disease increases. For example, knockdown of alleles pushing toward the female-favored trait value could cause expression that goes beyond the male-favored trait value, leading to pathologies of extreme maleness. Perturbations in the other direction may cause pathologies of extreme femaleness. 

\section{Immune triggers and suppressors}

Hosts must trigger immunity rapidly in response to pathogen invasion. The need for speed likely causes some false alarms. Unnecessary immune expression is costly, demanding strong negative regulators to shut down false responses. Immunity depends on strong rapid triggers and strongly opposed negative regulators \autocite{frank19evolution}.

Diverse, ancient families of immune triggers span the tree of life. Examples include peptidoglycan recognition proteins, gram-negative binding proteins, Toll-like receptors, nucleotide-binding domain leucine-rich repeats (NLRs), and the cyclic GMP-AMP synthase (cGAS) system that detects double-stranded DNA (dsDNA). 

Widespread negative regulators often oppose the triggers. Post-transcriptional modification of innate sensors and downstream molecules suppresses the effect of the pattern recognition receptor triggers. The cascade triggered by the dsDNA sensor cGAS has several negative regulators. Numerous noncoding micro-RNAs regulate immunity, including repression of the NLR triggers and their immune cascade.

Imbalances in strongly opposed triggers and suppressors can lead to disease. For example, some animals trigger type I interferon (INF) to start an immune response. Negative regulators of this trigger control immunity. Various immune-related disorders associate with type I IFN misexpression \autocite{crow22the-type}.

The quick trigger against potentially foreign DNA by cGAS is susceptible to autoimmunity. In the cGAS response, misregulation and immune-related disease arise in spite of the many negative regulators of the cGAS cascade \autocite{ablasser19cgas}.

\section{Behavioral triggers and suppressors}

A similar opposition of speedy triggers and slower suppressors arises in fear regulation. Potential danger requires quick action, often mediated by fear. False triggers require strongly opposed negative regulators.

For example, the brain's insular cortex responds to cues of fear \autocite{klein21fear}. Negative bodily feedback such as heart rate deceleration opposes the fearful state. Experimental perturbation of body-brain communication disrupts the balance between fear promoters and suppressors. This opposition of strong regulatory forces may explain the insular cortex's role in anxiety and addiction disorders and hyperactivity.

Amygdala-based fear responses also have various negative regulators \autocite{moscarello22the-central}. Imbalance associates with psychiatric disorders \autocite{limoges22dynorphin/kappa}.

Dopamine expression typically associates with positive situations, the opposite of fear stimulating triggers. Interestingly, dopamine expression is partly controlled by slow stimulants that are opposed by fast inhibitors \autocite{soden23circuit}. In particular, neurons within the lateral hypothalamus produce the slow-acting stimulatory neuropeptide neurotensin which is opposed by the fast-acting inhibitory neurotransmitter GABA (3-aminobutyric acid).

Perhaps a slowly developing beneficial situation favors dopamine stimulation. If that situation suddenly goes bad, a rapid behavioral reversal is needed, creating the opposition of strong forces on dopamine regulation and significant potential for misregulation. Dopamine misregulation associates with various disorders.

\section{Conclusion}

The evolutionary tuning of strongly opposed forces is particularly challenging. For example, the intensity of pathogen attack or the frequency of true danger varies over time and space. Greater intensity favors a lower threshold for expression of a quick trigger and a higher threshold for expression of a negative regulator.

Mating between parents from populations with different environments may lead to a mismatch of opposing forces and a greater potential for disease. Broader divergence between populations may lead to hybrid incompatibility.

\reviset{For speed-accuracy tradeoffs, I do not know of prior studies about parental divergence and offspring decay. For genetic conflicts, the potential association of genetic distance between parents and offspring degradation has been widely discussed \autocite{frank89the-evolutionary,frank91divergence,hurst91causes,johnson10hybrid}.}

\reviset{In humans, this theory predicts a positive association of the genetic distance between parents and the amount of disease in offspring for those particular traits that are shaped by strongly opposing forces. Current genetic methods can infer ancestry. This prediction may be testable in clinical settings with large samples of patients who have identifiable disorders in the focal traits.}

\section*{Acknowledgments}

\noindent The Donald Bren Foundation, National Science Foundation grant DEB-1939423, and DoD grant W911NF2010227 support my research. 


\mybiblio	


\end{document}